\documentclass[lettersize,journal]{IEEEtran}
\usepackage{amsmath,amsfonts}
\usepackage{algorithmic}
\usepackage{algorithm}
\usepackage{array}
\usepackage[caption=false,font=normalsize,labelfont=sf,textfont=sf]{subfig}
\usepackage{textcomp}
\usepackage{stfloats}
\usepackage{url}
\usepackage{bm}
\usepackage{verbatim}
\usepackage{physics}
\usepackage{graphicx}
\usepackage{listings}
\usepackage{courier}
\usepackage[inline]{enumitem}
\usepackage{cleveref}
\usepackage{xspace}
\usepackage{booktabs}
\usepackage[OT1]{fontenc}
\usepackage[style=ieee]{biblatex}
\usepackage{soul}

\Crefformat{figure}{#2Fig.~#1#3}

\newcommand*{\Ybus}[1]{$\mathbf Y_{\text {bus}}$}
\newcommand*{\Ibus}[1]{$\mathbf I_{\text {bus}}$}

\begin{document}

\title{Bus Admittance Matrix Revisited: Is It Outdated on Modern Computers?}

\author{Hantao Cui,~\IEEEmembership{Senior Member,~IEEE}
\thanks{This work is supported by U.S. National Science Foundation, Award
\#2226826.}%
}

\markboth{Preprint}%
{Shell \MakeLowercase{\textit{et al.}}: Bus Admittance Matrix Revisited}

\maketitle

\begin{abstract}

Bus admittance matrix is widely used in power engineering for modeling networks.
Being highly sparse, it requires fewer CPU operations when used for calculations.
Meanwhile, sparse matrix calculations involve numerous indexing and scalar operations,
which are unfavorable to modern processors.
Without using the admittance matrix, nodal power injections and the corresponding
sparse Jacobian can be computed by an element-wise method, which consists of a highly regular,
vectorized evaluation step and a reduction step.
This paper revisits the admittance matrix from the computational performance perspective by comparing
it with the element-wise method.
Case studies show that the admittance matrix method is generally slower
than the element-wise method for grid test cases with thousands to hundreds of thousands
of buses, especially on CPUs with support for wide vector instructions.
This paper also analyzes the impact of the width of vector instructions and memory speed
to predict the trend for future computers.

\end{abstract}

\begin{IEEEkeywords}
Bus admittance matrix, sparse matrix, high-performance computing, vectorization,
single-instruction multiple data (SIMD).
\end{IEEEkeywords}

\section{Introduction}
\IEEEPARstart{B}{us} admittance matrix (also known as admittance matrix,
$\mathbf Y$ matrix, or \Ybus{}) is ubiquitous in power engineering for network modeling \cite{tinneyPowerFlowSolution1967}.
The bus admittance matrix reduces a power grid with
$n_b$ buses and $n_l$ lines into an $n_b$-port network by relating bus current
injections to bus voltages. In realistic systems, a bus is only connected
to a few other buses, thus the admittance matrix is highly sparse. By adopting
sparse matrix techniques, calculations based on \Ybus{} requires significantly less
memory and arithmetic operations, making it superior to the then-popular
impedance matrix method \cite{tinneyCompensationMethodsNetwork1972}.

In many power systems studies such as power flow and stability simulations,
\Ybus{} is used to calculate bus power injections, as well as the Jacobian
matrix that contains the partial derivatives of injections concerning voltages and
angles. The computational performance to obtain network injections and Jacobians
is highly relevant because their computation time is only second to solving sparse
linear equations. Over the past decades, the central processing units (CPUs) in
computers have seen significant upgrades and shifted growth paradigms \cite{rossWhyCPUFrequency2008},
but not all computations can benefit equally. Calculations using \Ybus{} require sparse matrix-vector products,
where both the sparse matrix and vector are composed of complex numbers. Such
calculations are non-atomic to processors, thus the performance gain depends on
multiple factors and remains to be investigated.

Various techniques have been developed for modern CPUs with the primary goal to
reduce latency \cite{kirkProgrammingMassivelyParallel2017, etiemble45yearCPUEvolution2018}.
In the early 1980s, arithmetic calculations were already faster
than accessing data in memory, thus caches were added to CPU chips.
Instruction pipelining was later added to execute multiple, similar, and
independent instructions in different stages of completion. In the 90s, branch
prediction and out-of-order execution were added for relaxing dependent calculation.
With a probability of success, dependent calculations can be executed in
parallel. Another relevant technique is vector instruction known as single-instruction multiple data (SIMD), which allows
the processing of data in packs. To benefit from these techniques and achieve high performance,
the rule of thumb is to design algorithms and data structures that exhibit regularity \cite{flynnParallelArchitectures1996},
which translates to predictability for compilers and the hardware.

The sparse \Ybus{} method is not unique for network injection calculation.
An alternative method is termed the ``element-wise method" (or the vectorized method),
which calculates the element-wise power injections to form
branch-power vectors. Power injections into the line terminal are then summed up at the
connected buses to obtain the bus-wise power injections
 \cite{cuiHybridSymbolicNumericFramework2021,milanoPowerSystemModelling2010}. The equations for the
element-wise method are not unfamiliar, but the
method is often dismissed because typical power grids have more lines than
buses. Nevertheless, calculations in the element-wise method exhibit
repetitiveness, namely, the same calculations on different data, which may be
favored by modern CPUs.

This paper aims to provide a high-performance computing perspective on the \Ybus{}
method and the element-wise method for \textit{a}) bus power injection calculation and \textit{b})
sparse Jacobian matrix formulation. The method formulations, optimized implementations,
data access patterns, and computational complexity are presented. While this
work does not study the solution of equations, the formulated equations and
sparse Jacobians can be readily interfaced with efficient solvers. Rather, the
objective is to identify the method with higher performance and understand its
scalability limitations on modern CPUs.

The main contribution of this paper is the finding that the \Ybus{} method is
generally slower than the element-wise method for large-scale power systems.
This finding is based on rigorous implementation and benchmarks of highly optimized
implementations. All benchmarks performed on recent Intel and Apple
platforms support the observation, using test systems ranging from 14 to 82,000 buses and more.
Also, the reasons why the \Ybus{} method is trailing are discussed.
Further, the impact of CPU features and memory performance are
analyzed to assess the scalability promise of the two methods.

This paper is organized as follows: Section II discusses the fundamental
formulations of the \Ybus{} method and element-wise method.
Section III briefly introduces CPU features and computer
architecture relevant for result interpretation.
Section IV discusses data structures, computation steps, and implementation techniques for high performance.
Section V presents the benchmark data to compare different CPU generations across different
architectures. Section VI concludes the finding and predicts the scalability of
two methods on future hardware.

\section{Method Formulations}

This section discusses the formulations to calculate network power injections
and Jacobians.

\subsection{\Ybus{} Method} The complex power injections are calculated by
\begin{equation}
    \label{eq:Ybus_g_eqn}
    \bm{S} = \bm{V} \cdot \left(\vb{Y}_{\text bus} \bm{V} \right)^{*}
\end{equation}
where $\bm{S}$ and $\bm{V}$ are two complex-number vectors, respectively, for
power injections and voltage phasors.

The corresponding Jacobian matrix is given by
\begin{equation}
    \label{eq:Ybus_gy_matrix_concat}
    \bm{J} =
    \begin{bmatrix}
    \bm{J}_{11} & \bm{J}_{12} \\
    \bm{J}_{21} & \bm{J}_{22}
\end{bmatrix} =
    \begin{bmatrix}
    \Re ({\partial \bm S} / {\partial \bm \theta}) & \Re({\partial \bm S} / {\partial \bm V})\\
    \Im({\partial \bm S} / {\partial \bm \theta}) & \Im({\partial \bm S} / {\partial \bm V})
\end{bmatrix}
\end{equation}
where the derivative sub-matrices are calculated by \cite{milanoOpenSourcePower2005}
\begin{equation}
    \label{eq:Ybus_dS_dVa_matrix}
    \frac{\partial \bm S}{\partial \bm \theta} =
    j \cdot \text{diag}(\bm{V}) \left[ \text{diag}(\bm{I}_c^*) - (\vb{Y}_{\text{bus}} \text{diag}(\bm V))^* \right]
\end{equation}
\begin{equation}
    \label{eq:Ybus_dS_dVm_matrix}
    \frac{\partial \bm S}{\partial \bm V_m } =
    \text{diag}(\bm{I}_c^* \cdot e^{j \bm \theta}) + \text{diag}(\bm{V}) (\vb{Y}_{\text{bus}} \text{diag}(e^{j \bm \theta}))^*
\end{equation}

Although \eqref{eq:Ybus_dS_dVa_matrix} and \eqref{eq:Ybus_dS_dVm_matrix} are
compact, they are not of top efficiency as both require multiple sparse-sparse
multiplications. Section III will discuss an efficient two-pass algorithm to
simultaneously compute the two derivative matrices.

\subsection{Element-wise Method}

Consider a transmission line or a two-winding transformer whose off-nominal tap
ratio and phase shift $m e^{j\phi}$ is applied at the primary side. The complex
power injections $S$ into the two terminals, $h$ and $k$, are given by
\begin{equation}
    \label{eq:Ewise-Sh}
    S_{hk} = v_h^2 \frac{(y_h + y_{hk})^*}{m^2} - v_h v_k \frac{y_{hk}^*}{m} e^{j(\theta_h - \theta_k - \phi)}
\end{equation}
\begin{equation}
    \label{eq:Ewise-Sk}
    S_{kh} = - v_h v_k \frac{y_{hk}^*}{m} e^{-j(\theta_h - \theta_k - \phi)} + v_k^2 (y_k + y_{hk})^*
\end{equation}
where $y_h$ are the shunt admittance (including transformer core losses and magnetizing reactance) at the $h$ terminal, $y_k$ is the shunt admittance at the $k$ terminal, $y_{hk}$ is the series admittance, $v_h$ and $v_k$ are the voltage magnitudes, and $\theta_h$ and $\theta_k$ are the voltage phase angles of buses $h$ and $k$.

The power injection into bus $i$ can be obtained by adding up all the injections
from connected lines:
\begin{equation}
    \label{eq:Ewise-sum-up}
    S_{i} = \sum_{t=1}^{n_b} S_{it} \, ,  (i, t) \in L
\end{equation}
where $t$ is the other terminal of an existing line that connects buses $i$ and $t$, and $L$ is the set of lines denoted by bus pairs.
Each of \eqref{eq:Ewise-Sh} and \eqref{eq:Ewise-Sk} needs to be evaluated for $n_l$
times, and \eqref{eq:Ewise-sum-up} requires up to $2n_l$ additions.

To obtain the sparse Jacobian matrix, one first calculates the non-zero Jacobian entries and then
assemble them into a sparse matrix. The four power injection equations ($P_h$,
$Q_h$, $P_k$ and $Q_k$) can be obtained from \eqref{eq:Ewise-Sh} and
\eqref{eq:Ewise-Sk}. Taking the partial derivative of the four equations with respect
to four variables ($v_h$, $\theta_h$, $v_k$, and $\theta_k$) yields 16
expressions. All 16 expressions need to be evaluated for all $n_l$ lines to
obtain the complete Jacobian elements. Given a large number of Jacobian
expressions and the number of lines, this method is perceived as having high computational complexity.

\section{Modern CPU Features}

This section will walk through the features of modern CPUs to provide a
fundamental understanding of comparing the two methods. These features include
\textit{a}) cache and prefetch, \textit{b}) single-instruction multiple-data (SIMD) vectorization,
\textit{c}) pipelining and branch prediction. All these features serve one goal: to
reduce the overall latency.

Modern CPUs have multiple levels of on-chip cache. These caches are made of
static random-access memory (SRAM) so is faster to access than memory, which is
made of dynamic random-access memory (DRAM). For example, an Intel Core
i9-12900K has three levels of cache, namely, L1, L2, and L3, with
respective sizes of 80 KB (per core), 1.25 MB (per core), and 30 MB (total). The time to
access the three levels of cache is around 1 ns, 4 ns, and 40 ns, compared
with a typical memory access time of 80 ns. If the requested data is in the
cache, the access is a ``cache hit''. Otherwise, it is a ``cache miss'' and
will incur time to access the memory.

To utilize the on-chip cache and improve the cache hit rate, ``prefetch'' is
used to preemptively load data from memory to cache. Prefetch is accomplished by
both the CPU hardware and the compiler, which may insert prefetch instructions
into the machine code. Generally, data prefetch for an algorithm will have a higher
success rate if the algorithm access data in a more regular manner, such as
visiting a memory block linearly. Conversely, if the data locations can only be
determined at run time, prefetch success will likely decrease.

\begin{figure}[!t]
\centering
\includegraphics[width=\columnwidth, trim={.1cm 0 0 0}, clip]{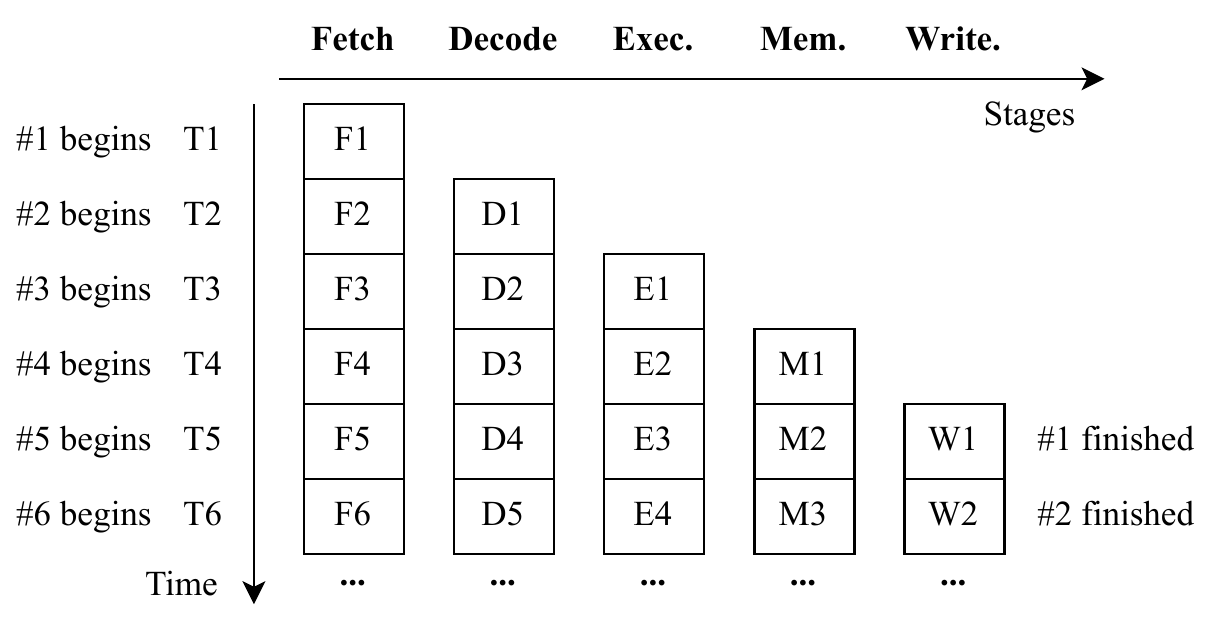}
\caption{Illustration of pipelining six instructions in a five-stage pipeline.}
\label{fig:pipelining}
\end{figure}

To reduce the number of instruction cycles for the same computations, a
single-instruction multiple-data (SIMD) instruction set is introduced. SIMD
instructions in the x86 platform include the Advanced Vector eXtension (AVX)
instructions, which can perform simultaneous arithmetic operations on four to
eight double-precision floating points, using AVX2 and AVX512, respectively. To
utilize SIMD, data operands need to be in the cache and ready to be loaded into
vector registers. On the other hand, using SIMD will add overhead to 
calculations that are done on scalars with mixed memory access.
In such cases, compilers will avoid generating SIMD instructions but instead 
use scalar operations.

Another relevant technique is instruction pipelining. Multiple stages are needed
to execute an instruction, including fetch, decode, execute, memory access, and
register write-back. Each stage is performed by a part of the CPU. Pipelining
aims to keep all parts busy while executing multiple instructions.
\Cref{fig:pipelining} illustrates the execution of six instructions in a
five-stage pipeline. The first instruction takes T1 to T5 to complete. Starting
from T6, each instruction will only take one time unit.

Although pipelines are efficient once warmed up, there is a downside. If one
instruction needs to be changed within the pipeline, the pipeline will stall and
needs to be flushed. Programs with \texttt{if-else} statements, for example,
will likely introduce instruction changes due to jumping. Modern x86 processors
have long pipelines, so for remediation, a branch prediction mechanism is
introduced. If the prediction is correct, the pipeline can continue, and only a
few cycles will be wasted. Otherwise, the time penalty will be much greater.

In summary, improving the cache hit rate, utilizing SIMD as much as possible,
and reducing instruction changes will
improve the performance of a program. As alluded to before, algorithms with a
regular data storage and access pattern will improve prefetch and improve the
cache hit rate. Also, programs with fewer conditional jumps will have better
pipeline performance. These metrics will be utilized to qualitatively assess the
efficiency of the \Ybus{} and element-wise method.

\section{Data Structure, Steps, and Implementations}

This section presents sufficient details on the data structure, computation steps, and implementations
that are crucial for a rigorous assessment of the two methods.

\subsection{Data Structure}

\subsubsection{\Ybus{} method}

The prominent data structure for storing \Ybus{} is the sparse matrix technique,
which is comprehensively covered in the literature
\cite{chowPowerSystemModeling2020,gloverPowerSystemAnalysis2022}.
The most common sparse formats to store \Ybus{} are Compressed Sparse
Column (CSC) and Compressed Sparse Row (CSR), and a trade-off exists between them.
The CSR format is faster than CSC for sparse matrix-vector product
operation but slower for factorization. If \Ybus{} is in CSR, the resultant
$\bm{J}$ will be in CSR by default. Then, $\bm J$ will need to be converted to
CSC every time to interface with fast solvers like KLU \cite{davisUserGuideKLU2022}.
As a result, it is more common to store \Ybus{} and $\bm J$ in the CSC format.

\subsubsection{Element-wise method}

The element-wise method does not require special data structures besides the
array. To facilitate SIMD and pipelining, the same parameter or variable of all
device instances is stored in a one-dimensional array. For example, an array
$\bm{v}_h$ will hold the sending-end voltage magnitudes of all lines. The data
structure guarantees that all elements of $\bm{v}_h$ are in contiguous memory.

The contiguous-memory requirement is also applied to the arrays of complex
numbers. That is, the real and imaginary parts of complex numbers are stored in
two separate arrays with an element type of floating point numbers.
Also, one can separate the
real power from reactive power in \eqref{eq:Ewise-Sh} and \eqref{eq:Ewise-Sk} so
that no complex-valued arrays are needed.

\subsection{\Ybus{} Method: Computation Steps and Algorithms}

\subsubsection{Power injection computation}

The following steps are performed in order to calculate the power injections:
\begin{enumerate} %
    \item Update the complex arrays $\bm U = e^{j \bm \theta}$ and $\bm V = \bm{V}_m \cdot \bm U$
    \item Calculate sparse matrix-vector product $\bm{I}_{\text bus} =
    \bm{Y}_{\text bus} \bm{V}$
    \item Calculate two vector product $\bm{V} \cdot \bm{I}^*_{\text bus}$
\end{enumerate}
where $\bm \theta$ is the array of voltage phase angles, $\bm{V}_m$ is the array
of voltage magnitudes, and $\bm I_{bus}$ is the array of current injections.

Steps 2 and 3 favor different data structures for $\bm{V}$ for performance. For
Step 2, the access into $\bm{V}$ is random, which depends on the current column
index of the non-zero element in \Ybus{}. Placing together the real and
imaginary parts of each complex element in $\bm{V}$ will enable the retrieval of
two values in one access. For Step 3, storing the real and imaginary parts of
$\bm{V}$ and \Ibus{} separately and linearly will enable SIMD support
with improved pipeline efficiency. This conflict is fundamental
and can only be alleviated by copying memory, which will incur overhead.

\subsubsection{Jacobian calculation}
A two-pass algorithm \cite{schaferEfficientOpensourceImplementation2018} is
adopted to compute $\partial \bm{S} / \partial \bm{\theta}$ and $\partial \bm{S}
/ \partial \bm{V_m }$ simultaneously and reuse intermediate values. This method
observes the sparsity patterns of $\bm{J}$ and \Ybus{}. That is, calculations in
\eqref{eq:Ybus_dS_dVa_matrix} and \eqref{eq:Ybus_dS_dVm_matrix} only involves
the product of \Ybus{} by diagonal matrices and the addition (or subtraction) of
diagonal matrices. Since these operations will not introduce non-zero elements
at locations where the corresponding \Ybus{} element is zero, the Jacobians
$d\bm{S} /d\bm{\theta}$ and $d\bm{S} /d \bm{V}_m$ will have the same sparsity
pattern as \Ybus{}. More precisely, the locations of nonzeros in the Jacobians
are subsets of the nonzeros in \Ybus{} because some values may become zeros.
Nonetheless, these small numbers of zeros can be stored to reuse the sparsity
pattern of \Ybus{}.

\Cref{alg:two-pass-csc} adopted from \cite{veraSanPenGridCal2022} presents the
pseudo-code for the two-pass method for CSC sparse matrix. In the inputs,
$\bm{Y}_p$, $\bm{Y}_i$ and $\bm{Y}_v$ are the column pointers, row indices, and
nonzeros of the CSC storage of \Ybus{}, \Ibus{} is the array for bus current
injections. In the initialization phase, copying the non-zero values of \Ybus{}
to that of $d\bm{S} /d\bm{\theta}$ and $d\bm{S} /d \bm{V}_m$ only involve linear
memory access and is thus rapid. The first pass computes the \Ybus{}-vector
product in \eqref{eq:Ybus_dS_dVa_matrix} and \eqref{eq:Ybus_dS_dVm_matrix}, and
the second pass computes the remainder.

\begin{algorithm}[t]
\caption{Two-pass method for Jacobian using CSC}
\begin{algorithmic}[1]
\STATE {\textsc{Input}:} $\bm Y_p$, $\bm Y_i$, $\bm Y_v$, $d\bm{S}_\theta$,
$d\bm{S}_V$, \Ibus{}, $\bm{V}$, $\bm{U}$ \STATE {\textsc{Initialize}:} \STATE
\hspace{0.5cm}$\bm{I}_{bus} = 0$, $d\bm{S}_\theta.{\text {nzval}} \mathrel{.}=
\bm {Y}_v$, $d\bm{S}_V.{\text {nzval}} \mathrel{.}= \bm {Y}_v$ \STATE
{\textsc{Pass 1}:} for $j=1:n_b$  ($j$ is column index) \STATE \hspace{0.5cm}
for $k=\bm{Y}_p[j]:(\bm{Y}_p[j+1]-1)$  ($k$ is $j$'s index range) \STATE
\hspace{1cm} $\bm{I}_{bus}[\bm{Y}_i[k]] \mathrel{+}= \bm{Y}_v[k] * \bm{V}[j]$
\STATE \hspace{1cm} $d\bm{S}_\theta.\text{nzval}[k] \mathrel{\times}= \bm{V}[j]$
\STATE \hspace{1cm} $d\bm{S}_V.\text{nzval}[k] \mathrel{\times}= \bm{U}[j]$
\STATE \hspace{0.5cm} end for $k$ \STATE end for $j$ \STATE {\textsc{Pass 2}:}
for $j=1:n_b$
\STATE \hspace{0.5cm} for $k=\bm{Y}_p[j]:(\bm{Y}_p[j+1]-1)$\textbf{} \STATE
\hspace{1cm} $i = \bm{Y}_i[k]$ ($i$ is element $k$'s row number) \STATE
\hspace{1cm} $d\bm{S}_V.\text{nzval}[k] = \bm{V}[i]\times (d\bm{S}_V[K])^*$
\STATE \hspace{1cm} if $i == j$ \STATE \hspace{1.5cm}
$d\bm{S}_\theta.\text{nzval}[k] \mathrel{-}= \bm{I}_{\text bus}[j]$ \STATE
\hspace{1.5cm} $d\bm{S}_V.\text{nzval}[k] \mathrel{+}= (\bm{I}_{bus}[j])^*
\times \bm{U}[j]$ \STATE \hspace{1cm} end if \STATE \hspace{1cm}
$d\bm{S}_\theta.\text{nzval}[k] = (1im) \times
(d\bm{S}_\theta.\text{nzval}[k])^* \times \bm{V}[i]$ \STATE \hspace{0.5cm} end
for $k$ \STATE end for $j$ \STATE \textsc{Return}: $d\bm{S}_\theta$, $d\bm{S}_V$
\end{algorithmic}
\label{alg:two-pass-csc}
\end{algorithm}

The two-pass method is more efficient than directly performing sparse matrix multiplications
because
\begin{enumerate}
    \item Memory allocation for sparse matrices is avoided due to the reuse of
    sparsity pattern.
    \item Memory access to the non-zero values of $d\bm{S} /d\bm{\theta}$ and
    $d\bm{S} /d \bm{V}_m$ are linear, thus data are more likely to have been
    prefetched to cache.
\end{enumerate}

Also, note that a slightly more efficient version can be obtained for CSR-stored
\Ybus{}. Since many sparse linear solvers only support the CSC format, a
CSR-to-CSC conversion will be required at every step. The overhead of the
conversion may out-weight the gain from using the CSR format.

\subsubsection{Jacobian matrix assembly}
The $\bm{J}$ matrix with real values can be obtained by concatenation as given
in \eqref{eq:Ybus_gy_matrix_concat}, but sparse matrix concatenation is slow due
to index sorting and memory allocation every time two sub-matrices are
concatenated.

\Cref{alg:J_indexing} is proposed to facilitate rapid assembly of the real and
imaginary parts of the complex Jacobians into $\bm{J}$, which is of the size
$2n_b\times 2n_b$. With pre-allocated sparsity pattern for $\bm J$, the assembly
operation is essentially copying the derivative elements to proper locations in
$\bm J$. Since the sparsity patterns do not change, one can pre-calculate the
indexing arrays $\bm{P}_{xy}$ into $\bm J$'s non-zeros for corresponding
sub-matrices.

\Cref{alg:J_indexing} is generalized for cases where all non-zeros of
$\bm{J}_{xy}$ exist in $\bm{J}$ but not necessarily the other way. For a basic
power flow calculation, each sub-matrices will have identical sparsity pattern
to \Ybus{}. This may not be the case when series devices other than lines and
transformers (such as TCSC) exist. \Cref{alg:J_indexing} is thus designed to
obtain the indices based on matching row and column coordinates.

\begin{algorithm}[t]
\caption{Obtain indices to write to sparsity pattern}\label{alg:J_indexing}
\begin{algorithmic}[1]
\STATE {\textsc{Input}:} $\bm J_p$, $\bm J_i$, $\bm{Y}_{\text{row}}$,
$\bm{Y}_{\text{col}}$, $\bm{P}_{xy}$ where $x, y \in (1, 2)$ \STATE
{\textsc{Initialize}:} \STATE \hspace{0.5cm}$\bm{P}_{xy} \mathrel{.}= 0$ where
$x, y \in (1, 2)$ \STATE {\textsc{Loop for $P_{xy}$}:} for
$i=1:\text{length}(\bm{Y}_{\text{col}})$ \STATE \hspace{0.5cm} $c_{ol} =
Y_{col}[i] + (y-1) * n_b $ \STATE \hspace{0.5cm} $r_{ow} = Y_{row}[i] + (x-1) *
n_b $ \STATE \hspace{0.5cm} for $k=\bm{J}_p[c_{ol}]:(\bm{J}_p[c_{ol}+1]-1)$
\STATE \hspace{1cm} if $J_i[k] \mathrel{==} r_{ow}$ \STATE \hspace{1.5cm}
$P_{xy}[i] = k$; break; \STATE \hspace{1cm} end if \STATE \hspace{0.5cm} end for
$k$ \STATE end for $i$ \STATE \textsc{Run} the loop with $(x, y) = (1, 1), (1,
2), (2, 1), \text{and}~(2, 2)$ \STATE \textsc{Return}: $\bm{P}_{11}$,
$\bm{P}_{12}$, $\bm{P}_{21}$, $\bm{P}_{22}$
\end{algorithmic}
\end{algorithm}

Using the pre-calculated indices, the Jacobian submatrices can be rapidly copied
by \texttt{J.nzval[$P_{xy}$] .= $J_{xy}$.nzval}. This method is efficient
because
\begin{enumerate}
    \item no random access to the sparse matrix $\bm J$ is needed. Time is thus
    saved by avoiding index look-ups,
    \item the memory copy operations visit the sub-matrix non-zeros linearly,
    which improves cache performance.
\end{enumerate}

\subsection{Element-wise Method Computation Steps}

\subsubsection{Vectorized calculation}
The element-wise method does not require sophisticated algorithms for
calculation. \ul{The principle of the element-wise method is to separate
CPU-bound vectorized evaluations from memory-bound operations and maximize SIMD
utilization for the former}. Since all parameters and variables are stored in
arrays of an equal length, the evaluation of all expressions for the same device
can be packed in the same loop, in which the same index is used to access the
arrays. An illustration of the calculation loop is given in
\Cref{lst:g-elemwise-loop},
\begin{lstlisting}[caption={Illustration of element-wise power injection calculation. Part of the expressions is omitted for brevity.},captionpos=b,label=lst:g-elemwise-loop]
for m in 1:n_line
    Ph[m] = vh[m]*(vh[m]*itap2ghghk[m] - ...)
    Pk[m] = vk[m]*(vk[m]*ghk[m] - ...) ...
end
\end{lstlisting}
where all constants, such as \texttt{itap2ghghk=$(g_{h} + g_{hk})/T^2_{ap}$},
are calculated once ahead of time.

There are three benefits of element-wise evaluations:
\begin{enumerate}
    \item the same parameters and variables are used in multiple expressions. As
    a result, cached values may be reused.
    \item since the expressions across devices are homogeneous, the compiler can
    generate SIMD instructions to simultaneously evaluate one expression for multiple
    instances.
    \item no branch exists in the calculation code, so the execution pipeline
    can be sustained.
\end{enumerate}

After calculating the line-wise power injections, the bus-wise power injections
are calculated using \eqref{eq:Ewise-sum-up}. This step is termed ``reduction"
as it reduces the number of injections from $n_l$ to $n_b$. Given that multiple
lines can connect to the same bus, each reduction involves three operations:
load, add, and store. Also, for a given order of lines, their connected buses
can be random, thus random access is required for the reduction.

\subsubsection{Jacobian matrix assembly}
There are multiple ways to assemble the sparse Jacobian matrix from elements.
This step is also a ``reduction'' because multiple non-zero values for the same
location need to be reduced into one by summation. The quickest implementation
is to convert the coordinates-value triplets to a new sparse matrix. Due to
memory allocation and sorting, this method is usually the slowest.

\begin{figure}[!t]
\centering
\includegraphics[width=0.9\columnwidth, trim={.5cm 0 0 0}, clip]{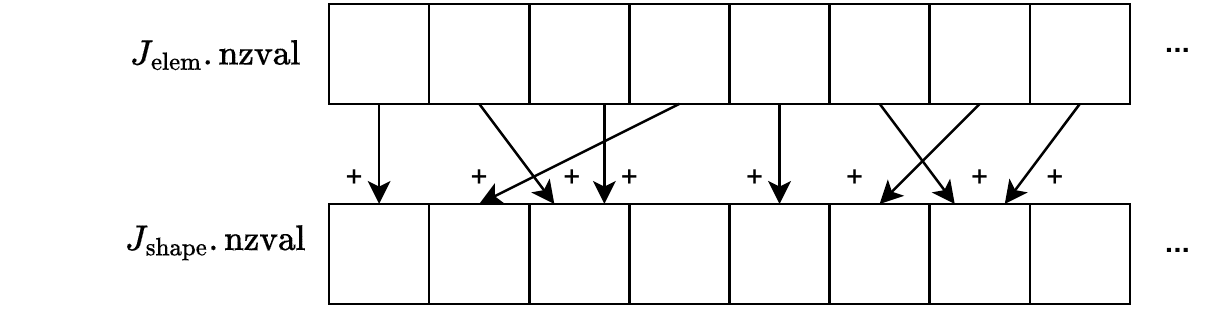}
\caption{Copy-adding non-zero elements to the sparsity pattern storage.}
\label{fig:standard-copy-add}
\end{figure}

Alternatively, one can pre-build the sparsity pattern matrix as in the \Ybus{}
method and copy-add elements to the non-zero array of the sparsity pattern
matrix. \Cref{fig:standard-copy-add} illustrates this process, where each
element in $\bm J_{\text{elem}}.{\text{nzval}}$ require four operations for an
element in $\bm J_{\text{shape}}.{\text{nzval}}$. These four operations include
two loads, one add, and one store.

\begin{figure}[!t]
\centering
\includegraphics[width=0.9\columnwidth, trim={.5cm 0 0 0}, clip]{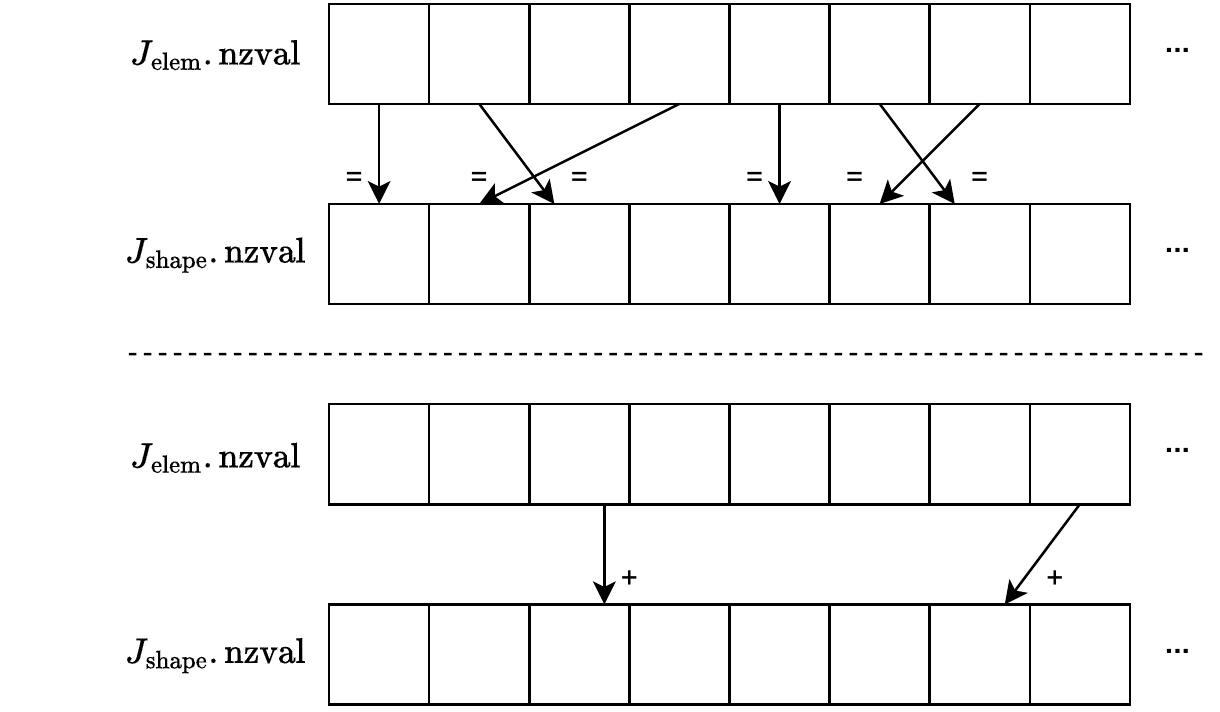}
\caption{Two-step technique with separate copying and copy-adding.}
\label{fig:two-step-copy-add}
\end{figure}

We propose a two-step implementation technique to separate the initial copy-only
operations from subsequent copy-add operations. The two steps are illustrated in
\Cref{fig:two-step-copy-add}.
The first step will copy elements from $\bm J_{\text{elem}}.{\text{nzval}}$ only
when the destinations in $\bm J_{\text{shape}}.{\text{nzval}}$ are visited for
the first time. It reduces the operations in the first step to one load and one
store. Subsequent copy-add operations will cover the remaining elements, which
will require four operations. This method eliminates one load and one add
operation for \texttt{length($\bm J_{\text{shape}}.{\text{nzval}}$)} times.
Given the reduced memory access, the total time will be considerably reduced.
Still, the reduction step has no parallelism and can be more time-consuming
than the equation evaluation.

\section{Case Studies}
This section presents comprehensive benchmark studies on the two methods for a
variety of power grid test cases, ranging from 14 to 82,000 buses and more
\cite{zimmermanMATPOWERSteadyStateOperations2011}. Micro-benchmarks are first
shown to demonstrate the performance gains by the techniques described in
Section IV. Next, the \Ybus{} and the element-wise methods are benchmarked for
grid test cases to identify the top performer, analyze the time breakdown, and
understand the limitations. Further, multiple x86 and ARM-based computer systems
are utilized to understand the trend of scalability as computer systems evolve.

All the discussed algorithms are implemented and optimized in Julia.
As a compiled language, Julia provides superior performance for code that is
optimally implemented. Generated machine code is inspected to guarantee SIMD
vectorization whenever possible.
All floating point numbers are stored in the 64-bit, IEEE 754-compliant format.
For complex numbers, the real and imaginary parts are stored using 64 bits.

\subsection{Micro-benchmarks of Computation Steps}
This section uses the Synthetic USA system with 82,000 buses on a computer with
Intel Core i9-12900K and 3600 MHz DDR4 RAM. The tested 12900K CPU only supports
AVX2, which is a 256-bit wide SIMD instruction.

\subsubsection{\Ybus{} method} The time to compute power injections is first
measured. Specifically, the power injections computation contains two steps: \textit{a})
phasor calculations and \textit{b}) sparse matrix and vector calculations, which
include steps 2 and 3 in Section IV.B. Cases are compared
with the real and imaginary parts of complex phasors stored interleaved or
separately. As shown in \Cref{tab:Ybus-interleave-separate}, separate storage
reduces the time to update the voltage vectors by using SIMD, although extra
time is required to change the data structure before the power calculations.
Note that the best time performance is given in the bold text in all subsequent tables.

\begin{table}[H]
\caption{P/Q calculation time in $\mu s$ for \Ybus{} with different storage}
\centering
\begin{tabular}{@{}lrrr@{}}
\toprule
           & $\bm V$ and $\bm U$ Calc. & Power Calc. & Total Time\\ \midrule Interleave & 557
& 722          & 1279  \\
Separate   & 137           & 806          & \textbf{943}   \\ \bottomrule
\end{tabular}
\label{tab:Ybus-interleave-separate}
\end{table}

The Jacobian computation has two steps: \textit{a}) $d\bm{S} /d\bm{\theta}$ and $d\bm{S}
/d \bm{V}_m$ calculation, and \textit{b}) assembly into $\bm J$. For Step \textit{a}), matrix
multiplication and the two-pass method are compared. For Step b), the
concatenation and in-place methods are compared. It can be seen that the
two-pass method is several times faster than the plain matrix multiplication,
and the in-place copy approach is nearly 25 times faster than concatenation. In
following comparisons with the element-wise method, the fastest methods will be
used.

\begin{table}[H]
\caption{Jacobian calculation and assembly time in $\mu s$ for \Ybus{} method}
\centering
\begin{tabular}{@{}cccc@{}}
\toprule
\multicolumn{2}{c}{Derivatives Calculation} & \multicolumn{2}{c}{Matrix
Assembly} \\ \midrule Matrix Multiplication         & Two-Pass        &
Concatenation      & In-Place Copy      \\
13,263              & \textbf{2,733}           & 15,350       & \textbf{602}           \\
\bottomrule
\end{tabular}
\label{tab:Ybus-jac-time}
\end{table}

\subsubsection{Element-wise method}

Both the power injection and Jacobian calculations in the element-wise method
has two steps: evaluation and reduction. For power injection calculation, with
and without SIMD are tested. \Cref{tab:elemwise-eqn-time} shows that with the
array-based data structure, the SIMD-enabled case only takes about 1/4 of the time
without SIMD. This is expected because the processor supports AVX2,
which packs four 64-bit floating-point calculations in one instruction.

\begin{table}[H]
\caption{Element-wise power injection computation time in  $\mu s$}
\centering
\begin{tabular}{@{}ccc@{}}
\toprule
\multicolumn{2}{c}{Evaluation Methods} & \multicolumn{1}{l}{Reduction}
\\\midrule No SIMD          & SIMD         &                               \\
1,179            & \textbf{299}          & \textbf{183}                           \\ \bottomrule
\end{tabular}
\label{tab:elemwise-eqn-time}
\end{table}

For the Jacobian computation, the evaluation time and the three reduction methods
are compared in \Cref{tab:elemwise-time-gy}. For the reduction part, creating a
new matrix is much more expensive than the copy-add and the two-step method.
Also, calculating all 1,665,936 Jacobian elements takes merely 848 $\mu s$,
compared with 1,830 $\mu s$ for the two-step reduction method. Therefore, the
bottleneck for Jacobian computation in the element-wise method is in the
reduction.

\begin{table}[H]
\caption{Element-wise Jacobian computation time $\mu s$}
\centering
\begin{tabular}{@{}cccc@{}}
\toprule
Vectorized Evaluation & \multicolumn{3}{c}{Reduction Methods} \\ \midrule & New matrix
            & Copy-add & Two-step \\
\textbf{848}         & 13,873  & 2,025     & \textbf{1,830}    \\ \bottomrule
\end{tabular}
\label{tab:elemwise-time-gy}
\end{table}

\subsubsection{Comparing \Ybus{} and element-wise methods}
The best computation times for both methods are given in
\Cref{tab:ybus-elemwise-compare}. It can be seen that the \Ybus{} method is
slower than the element-wise method for both power injection and Jacobian
calculations. Since the theoretical computational complexity is in the
same order of magnitude (with the \Ybus{} method having slightly less
calculations), the time difference needs to be investigated at the execution
level.

\begin{table}[H]
\centering
\caption{Comparison of best computation time in $\mu s$ }
\label{tab:ybus-elemwise-compare}
\begin{tabular}{@{}lrr@{}}
\toprule
                 & \Ybus{} Method & Element-wise Method \\ \midrule
Power Computation           & 943         & \textbf{482}                 \\
Jacobian Computation        & 3,335       & \textbf{2,678}               \\ \bottomrule
\end{tabular}
\end{table}

We use the CPU statistics by the Linux \texttt{perf}  monitoring tool to compare
the element-wise Jacobian evaluation and the two-pass method. Note that the
element-wise method has more complexity on paper due to the number of elements,
but the two-pass method has multiple jumps. The two functions are executed for
1,000 runs, and the CPU statistics are shown in \Cref{tab:cpu-statistics}. It
can be seen that the \Ybus{} method requires more cache access and has nearly
twice the cache miss rate than the element-wise method. Also, the \Ybus{} method
has significantly more branches and much higher branch mispredictions.
Therefore, the total number of instructions for the \Ybus{} method is higher,
resulting in more cycles and a longer execution time.

\begin{table}[H]
\centering
\caption{CPU statistics for Jacobian calculation}
\begin{tabular}{@{}lrr@{}}
\toprule
\multicolumn{1}{c}{} & \multicolumn{1}{c}{\Ybus{} J Calculation} &
\multicolumn{1}{c}{Element-wise J Calculation} \\ \midrule Cycles
& 14,327,078,176                         & 4,577,698,819
\\
Cache Access         & 828,580,410                            & 581,952,974
\\
Cache Misses         & 156,647,072                            & 60,248,850
\\
Branches             & 5,474,516,729                          & 19,537,258
\\
Branch Mispredict.      & 173,767,459                            & 2,538
\\
Instructions         & 50,308,631,851                         & 3,690,449,376
\\ \bottomrule
\end{tabular}
\label{tab:cpu-statistics}
\end{table}

\subsection{Benchmark for Multiple Test Cases}

This section continues to use Intel 12900K for benchmarking the two methods for
nine power grid test cases. Results are shown in
\Cref{tab:12900k-time-comparison}. For the element-wise method,
\begin{enumerate*} [label=\itshape\alph*\upshape)]
\item the Jacobian reduction is the bottleneck of all its computations.
\item the power equation evaluation scales nearly linearly to system size.
\end{enumerate*}
For the \Ybus{} method, the Jacobian calculation is the bottleneck.

Several observations are made by comparison:
\begin{enumerate}
    \item For small cases such as the 14-bus and the 118-bus systems, the
    \Ybus{} method is faster in both power injection and Jacobian calculations.
    This is because all data are likely in the L1 or L2 cache, and the
    computation demand is insufficient to sustain the pipeline.
    \item The effect of pipelining can be seen by comparing the power
    calculation time for the 118-bus and the 300-bus systems. The two
    systems are of modest sizes so data can mostly fit into L1 and L2 caches.
    As the number of buses increases by 2.5x,  the calculation time increases
    by a factor of 2.6 for the \Ybus{} method.
    For the element-wise method, the calculation time only increases by 2.1x due to
    improved pipelining for SIMD.
    \item The power computation is faster using the element-wise method than the
    \Ybus{} method for cases with greater than 300 buses.
    This is possibly due to cache misses in the \Ybus{} method when performing
    sparse matrix-vector multiplication.
    \item The Jacobian computation is faster using the element-wise method for
    systems of any size. This is mostly due to the branching and jumping that
    are unfriendly to pipelining, as discussed in Section V.A.3.

\end{enumerate}

Overall, the element-wise method is faster than \Ybus{} method for both power
injection and Jacobian computations for systems ranging from 300 to 82k buses.

\begin{table*}[]
\centering
\caption{Computation time for the \Ybus{} and the element-wise methods for
systems of different sizes. \\ Time is given in nanoseconds for rows 1 and 2 and
in microseconds for the remainder.}
\begin{tabular}{lrrrrrrrr}
                   & \multicolumn{4}{l}{\Ybus{} Method} &
\multicolumn{4}{l}{Element-wise Method}        \\ \toprule Case               &
Phasor Calc.   & Power Calc.  & J Calc.   & J Red.  & Power Calc. & Power Red. &
J Calc.  & J Red.   \\ \midrule case14 (ns)            & \textbf{29.86}          & \textbf{93.21}
& 333.37    & 88.93   & 59.84       & 105.69     & \textbf{65.46}    & \textbf{87.41}    \\
case118 (ns)           & \textbf{207.65}         & \textbf{519.15}       & 2,819.56  & 565.71  &
537.21      & 300.17     & \textbf{1,044.30} & \textbf{869.42}   \\
case300            & 0.51           & 1.37         & 7.19      & 1.64    & \textbf{1.18}
& \textbf{0.54}       & \textbf{2.58}     & \textbf{2.04}     \\
case1354pegase     & \textbf{2.28}           & \textbf{5.94}         & 31.82     & 8.05    & 5.66
& 2.65      & \textbf{12.51}    & \textbf{13.29}    \\
case2736sp         & 4.75           & 12.42        & 65.87     & 16.70   & \textbf{9.92}
& \textbf{4.56}       & \textbf{15.49}    & \textbf{23.61}    \\
case9241pegase     & 15.49          & 79.08        & 341.35    & 75.46   & \textbf{45.89}
& \textbf{26.72}      & \textbf{124.80}   & \textbf{248.02}   \\
case\_ACTIVSg25k   & 41.77          & 214.54       & 818.18    & 176.18  & \textbf{92.62}
& \textbf{49.70}      & \textbf{230.34}   & \textbf{491.11}   \\
case\_ACTIVSg70k   & 117.00         & 665.77       & 2,291.57  & 504.21  &
\textbf{255.39}      & \textbf{142.50}     & \textbf{686.66}   & \textbf{1,536.49} \\
case\_SyntheticUSA & 137.09         & 806.48       & 2,733.15  & 602.55  &
\textbf{299.40}      & \textbf{183.85}     & \textbf{847.68}   & \textbf{1,829.67} \\
\bottomrule
\end{tabular}
\label{tab:12900k-time-comparison}
\end{table*}

\subsection{Computational Complexity Due to Grid Size}

While the element-wise method leads in performance, it is observed that the
power and Jacobian calculations for the two methods slow down differently. That
is, as the system size increases, the average computational complexity increases by more
for some operations than others. We can measure a rough metric coined as
the ``computation time per bus`` for each step. As the grid size increases, cache
misses and memory access increase,
thus a longer computation time per bus is expected.

To characterize the relative computation performance across systems, we
calculate the computation-time-per-bus ratios of all systems to that of the
118-bus system. This ratio is termed the ``time-cost ratio``.
A time-cost ratio smaller than one means that an operation takes less time per bus to perform, as compared to the 118-bus system.
Ratios greater than 1 correspond to super-linear complexity (namely, unfavorable scalability),
indicating worsening performance as system size increases.
The larger the time-cost ratio, the worse the scalability is.

\begin{figure}[htp!]
\centerline{\includegraphics[width=\columnwidth]{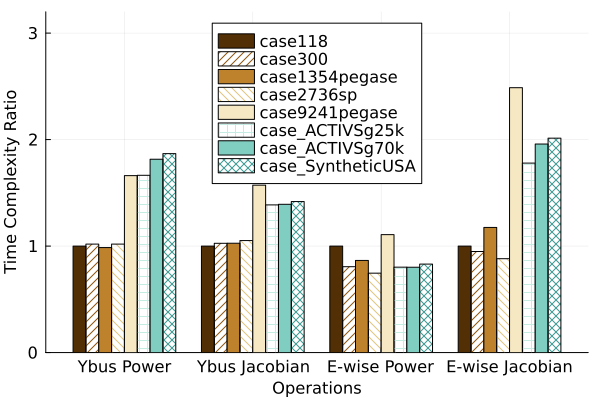}}
\caption{Time complexity ratio relative to the 118-bus system on 12900K}
\label{fig:time-ratio}
\end{figure}

Results are plotted in \Cref{fig:time-ratio}, which
shows that the \Ybus{} method has linear scalability up to the 2736-bus system
because data are likely in the CPU caches. The element-wise method for power
calculation has sub-linear scaling and is thus optimal compared with \Ybus{}.
Further, the Jacobian calculation in the element-wise method has worse
scalability than the \Ybus{} method. If this trend sustains, the \Ybus{} method
can be faster to compute the Jacobian than the element-wise method.

\subsection{Impact of SIMD Width}
As noted, the tested 12900K CPU only supports AVX2, which is 256-bit wide. It
is anticipated that CPUs with AVX512 support will be faster for the element-wise
evaluations. But is it the case across the board?

We benchmark the nine cases on an Intel 11900K CPU with 32GB DDR4 memory
clocking at 3200 MHz. Compared with 12900K, this CPU is one generation older in
architecture,  runs at a lower frequency, and has smaller caches. Also, the
memory is slower than that used in the 12900K system.
On the contrary, this CPU supports 512-bit wide SIMD through AVX512.
Comparing the results
shown in \Cref{tab:11900k} with \Cref{tab:12900k-time-comparison}, the following
observations are made:
\begin{enumerate}
    \item Phasor evaluation in the \Ybus{} method and the power evaluation in
    the element-wise method is faster for all systems due to AVX512 support.
    Given the system size, the operands can fit into the L2 cache. These
    evaluations are thus CPU-bound.
    \item In the \Ybus{} method, the power evaluation cannot utilize AVX512. Its
    computation performance is thus slower and proportional to the CPU clock frequency.
    \item In the element-wise method, the Jacobian evaluation is faster for
    small systems and slower for larger ones. The slowdown is due to memory
    access for large systems. Therefore, this evaluation switches from CPU-bound
    to memory-bound as system size increases.
\end{enumerate}

It can be concluded that AVX512 will reduce the computation time for
vectorized evaluations by a modest percentage, and it is most conducive to small
and medium-sized systems. For large systems, the whole computation becomes
memory-bound for both methods.

\begin{table*}[]
\centering
\caption{Benchmark on Intel i9-11900K with AVX512}
\begin{tabular}{@{}lrrrrrrrr@{}} & \multicolumn{4}{l}{Ybus Method}
                   & \multicolumn{4}{l}{Element-wise Method}       \\ \toprule
                   Case               & Phasor Calc. & Power Calc. & J Calc.  &
                   J Red. & Power Calc. & Power Red. & J Calc. & J Red.   \\
                   \midrule case14 (ns)            & \textbf{17.46}        & \textbf{123.93}
                   & 463.99   & 111.28 & 56.24       & 145.56     & \textbf{71.34}   &
                   \textbf{90.59}    \\
case118 (ns)           & \textbf{122.71}       & \textbf{605.35}      & 3,911.88 & 819.61 & 400.09
& 443.47     & \textbf{824.30}  & \textbf{1,027.40} \\
case300            & 0.30         & 1.66        & 10.06    & 2.20   & \textbf{0.72}
& \textbf{0.78}       & \textbf{1.93}    & \textbf{2.43}     \\
case1354pegase     & 1.38         & 6.89        & 43.85    & 9.63   & \textbf{3.39}
& \textbf{3.57}       & \textbf{10.86}   & \textbf{19.00}    \\
case2736sp         & 2.72         & 14.90       & 91.93    & 21.79  & \textbf{6.18}
& \textbf{6.31}       & \textbf{18.64}   & \textbf{32.99}    \\
case9241pegase     & 9.41         & 88.23       & 430.85   & 89.32  & \textbf{29.63}
& \textbf{34.76}      & \textbf{97.57}   & \textbf{255.11}   \\
case\_ACTIVSg25k   & 25.40        & 249.08      & 1,053.98 & 205.37 & \textbf{58.59}
& \textbf{65.77}      & \textbf{195.06}  & \textbf{394.81}   \\
case\_ACTIVSg70k   & 69.62        & 757.26      & 3,274.72 & 712.65 & \textbf{168.35}
& \textbf{185.77}     & \textbf{755.40}  & \textbf{2,041.78} \\
case\_SyntheticUSA & 81.58        & 969.50      & 3,956.52 & 875.01 & \textbf{205.78}
& \textbf{218.37}     & \textbf{959.30}  & \textbf{2,444.29} \\ \bottomrule
\end{tabular}
\label{tab:11900k}
\end{table*}

\begin{table*}[]
\centering
\caption{Benchmark on Apple M2}
\label{tab:apple-m2}
\begin{tabular}{@{}lrrrrrrrr@{}} & \multicolumn{4}{c}{Ybus Method}
                   & \multicolumn{4}{c}{Element-wise Method}        \\ \midrule
                   Case               & Phasor Calc. & Power Calc. & J Calc.  &
                   J Red. & Power Calc. & Power Red. & J Calc.  & J Red.   \\
                   \midrule case14 (ns)        & \textbf{42.17}        & \textbf{131.38}      &
                   360.44   & 100.81 & 92.32       & 184.30     & \textbf{93.86}    &
                   \textbf{140.80}   \\
case118 (ns)       & \textbf{281.18}       & \textbf{666.93}      & 2,916.67 & 761.69 & 823.96
& 529.02     & \textbf{688.47}   & \textbf{1,366.60} \\
case300            & \textbf{0.69}         & \textbf{1.49}        & 7.33     & 1.86   & 1.95
& 0.98       & \textbf{1.75}     & \textbf{3.47}     \\
case1354pegase     & \textbf{3.12}         & \textbf{7.38}        & 34.88    & 8.58   & 9.33
& 5.11       & \textbf{14.29}    & \textbf{23.33}    \\
case2736sp         & \textbf{6.29}         & \textbf{14.71}       & 77.46    & 17.67  & 16.42
& 7.79       & \textbf{18.17}    & \textbf{32.75}    \\
case9241pegase     & \textbf{21.13}        & \textbf{88.42}       & 355.29   & 69.13  & 95.33
& 48.83      & \textbf{124.08}   & \textbf{222.04}   \\
case\_ACTIVSg25k   & 57.13        & 227.29      & 883.04   & 157.33 & \textbf{190.25}
& \textbf{89.17}      & \textbf{229.67}   & \textbf{332.88}   \\
case\_ACTIVSg70k   & 160.08       & 645.92      & 2,592.08 & 452.63 & \textbf{521.21}
& \textbf{250.63}     & \textbf{1,025.92} & \textbf{962.33}   \\
case\_SyntheticUSA & 187.79       & 769.29      & 2,821.54 & 524.63 & \textbf{619.17}
& \textbf{296.88}     & \textbf{1,351.75} & \textbf{1,194.46} \\ \bottomrule
\end{tabular}
\end{table*}

\subsection{Impact of Improved Memory Speed}

At this point, it is understood that memory access is the bottleneck for large-scale system
calculations. Solutions from computer systems are to clock up the memory and
physically move the processor and memory closer. A notable system is the Apple
Silicon, which places the RAM beside the processor. This design has the benefit
of reduced memory latency, which may lead to performance gains for large-scale
systems. Understanding how this design impacts the performance of the two methods
will help us to predict their scalability.

\Cref{tab:apple-m2} shows the benchmark results on an Apple M2 system with 24 GB
LPDDR5 RAM running at 6400 MHz. Apple M2 has support for 128-bit SIMD, 192 KB L1 cache
, and 16 MB L2 cache. Comparing \Cref{tab:apple-m2} with
\Cref{tab:12900k-time-comparison} provides interesting observations:

\begin{enumerate}
    \item The phasor calculation using \Ybus{} and the power and
    Jacobian evaluation using the element-wise method on Apple M2 is consistently
    within 2x the time of Intel 12900K. This is because these computations are
    CPU-bound, in which the SIMD width plays a central role.
    \item The power evaluation in \Ybus{} is initially slower and later faster on M2
    than on 12900K. This is because, for small systems, the larger cache in 12900K can
    help reduce cache misses. For large systems, the problem will become memory-bound, thus M2 becomes faster.
    \item The Jacobian calculation in \Ybus{} is slower on M2 than on 12900K by a stable
    factor. This step cannot utilize SIMD and is impacted by both the CPU
    frequency and memory speed.
    \item The Jacobian reduction is initially CPU-bound but quickly becomes
    memory bound. Faster memory will bring major improvements to this step.
\end{enumerate}

\begin{figure}[htp!]
\centerline{\includegraphics[width=\columnwidth]{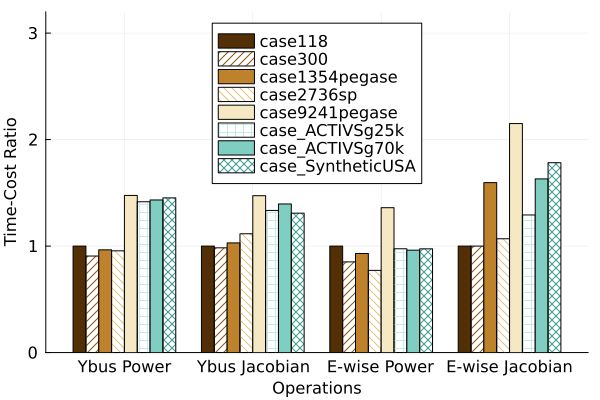}}
\caption{Computation time ratio relative to the 118-bus system on Apple M2}
\label{fig:time-ratio-m2}
\end{figure}

In addition, \Cref{fig:time-ratio-m2} shows the time-cost ratio relative to the
118-bus system on M2. It can be seen that for the 82k-bus Synthetic system, the
element-wise method scales slightly worse than the \Ybus{} method for Jacobian
calculation. This observation matches that on the Intel platform, although the
gap becomes smaller since faster memory is more helpful to the element-wise
method.

\begin{table*}[]
\centering
\caption{Computation time in microseconds using ultra-large grid test cases on Intel 12900K}
\label{tab:129k-ultra-large}
\begin{tabular}{@{}lrrrrrrrr@{}} & \multicolumn{4}{c}{Ybus Method}
     & \multicolumn{4}{c}{Element-wise Method}     \\ \midrule Case & Phasor
     Calc. & Power Calc. & J Calc. & J Red. & Power Calc. & Power Red. & J Calc.
     & J Red. \\ \midrule 164k & 275          & 1,721       & 5,806   & 1,478  &
     \textbf{608}         & \textbf{373}        & \textbf{2,084}   & \textbf{4,192}  \\
246k & 412          & 2,796       & 9,043   & 2,580  & \textbf{937}         & \textbf{583}
& \textbf{3,344}   & \textbf{7,369}  \\
328k & 549          & 4,060       & 12,511  & 3,938  & \textbf{1,346}       & \textbf{775}
& \textbf{4,577}   & \textbf{11,067} \\
410k & 689          & 5,208       & \textbf{15,948}  & \textbf{5,300}  & \textbf{1,915}      & \textbf{965}
& 6,008   & 16,740 \\ \bottomrule
\end{tabular}
\end{table*}

\begin{table*}[]
\centering
\caption{Computation time in microseconds using ultra-large grid test cases on Apple M2}
\label{tab:m2-ultra-large}
\begin{tabular}{@{}lrrrrrrrr@{}} & \multicolumn{4}{c}{Ybus Method}
     & \multicolumn{4}{c}{Element-wise Method}     \\ \midrule Case & Phasor
     Calc. & Power Calc. & J Calc. & J Red. & Power Calc. & Power Red. & J Calc.
     & J Red. \\ \midrule 164k & 367          & 1,698       & 5,911   & 1,158  &
     \textbf{1,256}       & \textbf{594}        & \textbf{2,906}   & \textbf{3,442}  \\
246k & 566          & 2,577       & \textbf{8,883}   & \textbf{1,776}  & \textbf{1,930}       & \textbf{912}
& 4,771   & 6,414  \\
328k & 757          & 3,435       & \textbf{12,079}  & \textbf{2,399}  & \textbf{2,619}       & \textbf{1,226}
& 6,953   & 9,590  \\
410k & 950          & 4,350       & \textbf{15,096}  & \textbf{3,006}  & \textbf{3,226}       & \textbf{1,538}
& 7,982   & 12,798 \\ \bottomrule
\end{tabular}
\end{table*}

\subsection{Outlook for Ultra-Large Systems}

The tested cases with up to 82k buses declare the element-wise method as a winner for
large-scale systems. Still, worse scalability is observed for the Jacobian
computation in the element-wise method. This indicates that, as the system size
further increases, the \Ybus{} method may take over. This section investigates
how the two methods will continue to scale for ultra-large cases to understand
the performance turning points.

The 82k-bus Synthetic USA system is repeated multiple times to create four
ultra-large systems. All buses and lines are duplicated by two to five times,
respectively, resulting in four cases. The performance of the two methods on
12900K and Apple M2 are reported in \Cref{tab:129k-ultra-large} and
\Cref{tab:m2-ultra-large}. It can be seen that the Jacobian reduction step in
the element-wise method continues to scale poorly. On the Intel platform, the
\Ybus{} method and the element-wise methods will be similarly slow when the
number of buses reaches about 328k. On the M2 platform, the tie happens around
246k buses.

Regarding the question that which of the two methods will be faster on future
systems, we argue that the element-wise method will continue to outperform the
\Ybus{} method for large-scale systems ranging from 1k to 410k buses or more. If
we take the best columns from \Cref{tab:129k-ultra-large} and
\Cref{tab:m2-ultra-large}, the element-wise method is a clear winner for power
calculation. For the Jacobian calculation, the best columns of the two methods will take roughly the same
time for the 410k bus system. Nevertheless, faster memory and AVX2/AVX512 will
moderately benefit the element-wise Jacobian calculation but not as much for
the \Ybus{} method. In addition, the Jacobian reduction method presented
in Section IV still has room for optimization, but that is beyond the scope of
this paper.

\section{Conclusions}

In this paper, we investigated the computational performance of the \Ybus{} method
for power injection calculation and sparse Jacobian formulation.
The main findings are as follows:

\begin{enumerate}
    \item The major issue with the \Ybus{} method is the sparse matrix-centered computations,
    which are not capable of utilizing SIMD and incurs many jumps due to indexing.
    \item Compared with the element-wise method, the \Ybus{} method is slower for practical
    systems with bus numbers ranging from thousands to a few hundred thousand on the x86 CPUs.
    \item The bottleneck for the element-wise method is reducing the Jacobian elements into a sparse matrix. The step scales poorly as the number of buses enters the realm of hundreds of thousands. This step is primarily constrained by memory speed.
\end{enumerate}

We conclude that the \Ybus{} method trails the element-wise method for practical
power grids on modern computers, especially those with wide SIMD support and fast memory.
Bus admittance matrix, as a network reduction method for increasing computation performance,
is missing the support for new processor features due to its computational irregularity.
Future work needs to investigate algorithms for reducing the time complexity of the Jacobian
reduction step in the element-wise method.

\printbibliography

\end{document}